\documentclass[sn-mathphys-num]{sn-jnl}

\usepackage[utf8]{inputenc}
\usepackage{color}
\usepackage{chngpage}
\usepackage{afterpage}
\usepackage[table]{xcolor}
\usepackage{natbib}
\usepackage{hyperref}
\usepackage{float}
\usepackage{soul}
\usepackage{caption}
\usepackage{subcaption}
\usepackage{graphicx}%
\usepackage{multirow}%
\usepackage{amsmath,amssymb,amsfonts}%
\usepackage{amsthm}%
\usepackage{mathrsfs}%
\usepackage[title]{appendix}%
\usepackage{textcomp}%
\usepackage{manyfoot}%
\usepackage{booktabs}%

\usepackage{todonotes,changes}
\usepackage{gensymb}
\usepackage{xurl}

\theoremstyle{thmstyleone}%
\theoremstyle{thmstyletwo}%
\theoremstyle{thmstylethree}%
\def\bea{\begin{eqnarray}}
\def\eea{\end{eqnarray}}
\def\la{\langle}
\def\ra{\rangle}

\begin{document}
\title{An entropy based comparative study of regional and seasonal distributions of particulate matter in Indian cities}

\author*[1]{\fnm{Suchismita} \sur{Banerjee}}\email{suchib.1993@gmail.com}

\author[1]{\fnm{Urna} \sur{Basu}}

\author[2]{\fnm{Banasri} \sur{Basu}}

\affil*[1]{\orgdiv{Department of Physics of Complex systems}, \orgname{S. N. Bose National Centre for Basic Sciences}, \orgaddress{\street{JD Block, Sector-III, Bidhannagar}, \city{Kolkata}, \postcode{700 106}, \state{West Bengal}, \country{India}}}

\affil[2]{\orgdiv{Interdisciplinary Statistical Research Unit}, \orgname{Indian Statistical Institute}, \orgaddress{\street{B T Road}, \city{Kolkata} \postcode{700 108}, \state{West Bengal}, \country{India}}}


\abstract{Particulate matter (PM), especially $\text{PM}_{2.5}$, is a critical air pollutant posing significant risks to human health and the environment in India. This study, using six years (2018-2024) of daily  $\text{PM}_{2.5}$ data, investigates the seasonal characteristics of the distributions of $\text{PM}_{2.5}$ concentrations across eleven Indian cities, selected from different regions of the country. We find that, while each city has its own unique seasonal patterns, all of them show a universal exponential decay in the tail of the $\text{PM}_{2.5}$ distribution for all the seasons. However, the decay rates of this tail vary across cities, highlighting regional and seasonal disparities in pollution levels. To quantitatively characterize the {\it randomness} of the seasonal $\text{PM}_{2.5}$ concentration distributions,  we compute Shannon entropy, a key information theoretic measure. This  allows for classifying  cities into different groups, according to the level of randomness observed  in their seasonal distributions. To further explore the inter-city relationships, we employ Jensen-Shannon divergence (JSD), a symmetric measure of relative entropy, to quantitatively assess the degree  of similarity in the $\text{PM}_{2.5}$ distributions among different cities. Remarkably, we find that several cities show very similar distributions in the winter months, which helps us to categories them into several groups. The groups obtained from these entropy based measures, namely, individual Shannon entropy and the JSD estimate, are consistent with each other, providing a robust framework for efficient air quality management and policy-making in India.}


\maketitle

\section{Introduction}
In recent years, concerns about air pollution have intensified worldwide, driven by mounting evidence of its detrimental impacts on human health~\cite{Bala_2019}, agriculture~\cite{Avne_2013,Gao_2020}, and the economy~\cite{Pand_2021}.  Despite making progress towards cleaner air, many  cities in the world  still experience episodes of air pollution with some urban centers frequently going beyond  the air quality levels permitted by the World Health Organization's (WHO) air quality guidelines.  In particular, among many key pollutants,  Particulate Matter (PM) is  a  harmful pollutant  posing  significant health risk.

PM is damaging on multiple scales, from localized pollution toxicity to large-scale impacts on climate change, making it a significant factor in geophysical and environmental challenges~\cite{Jac_2000}. PM is generally classified by size and origin, as well as by its chemical and physical characteristics. The two primary categories are $\text{PM}_{2.5}$ (particles with a diameter less than 2.5 $\mu m$, also known as `fine mode') and $\text{PM}_{10}$ (particles with a diameter less than 10 $\mu m$, or `coarse mode'). Fine particles can penetrate deeply into the lungs posing higher risks of pulmonary and respiratory diseases~\cite{Wang_2002,Harri_2000}
compared to their coarse counterparts. 
The dynamic nature of $\text{PM}_{2.5}$ results in both direct and indirect impacts on meteorological conditions, contributing to changes in radiation budgets, circulation patterns, and cloud albedo, which in turn exacerbate climate change~\cite{Wat_2002,Ram_2001}. 
The dual impact on public health and climate underscores the urgent need to understand the distributional behavior of $\text{PM}_{2.5}$, so that the mitigation of its emission can be done in an effective way.

India is a major hotspot for particulate matter production~\cite{Ham_2020,Don_2021,GUTTIKUNDA2014501,Pant_2019,Narayanan_2022,Watson_2021}. 
In recent years, rapid industrialization, urbanization, vehicular emissions, inadequate waste management, and an accelerating population growth have all contributed to the rise in PM levels across the country~\cite{Desh_2013}. Anthropogenic activities, notably the burning of agricultural residues, fossil fuels, and biomass, are primary contributors to fine particle emissions across the Indian subcontinent~\cite{Ram_2008,Venkatraman_2018,source}.
Recent studies have explored the sources, transport mechanisms, and concentrations of $\text{PM}_{2.5}$ in major Indian urban centers, offering a deeper understanding of the complex and persistent challenge of air pollution in these densely populated regions~\cite{Pit_2011,Ren_2011}.

Given the serious risks associated with the increase in levels of $\text{PM}_{2.5}$ in India, it is essential to control this pollutant. To develop effective strategies, it is crucial to understand the spatiotemporal behavior and different characteristics of the distribution of $\text{PM}_{2.5}$ in different regions.
Recent research~\cite{Kumar, Anand} on particulate matter pollution in India has provided valuable perspectives on its spatiotemporal dynamics and health implications. Studies have explored~\cite{Bran_2024,Bran_2017,Sharma_2022} time variations and averages of PM concentrations, revealing alarming levels of fine particulate matter ($\text{PM}_{2.5}$) and coarse particulate matter ($\text{PM}_{10}$) across urban and rural areas. For instance, research focusing on rural southern India highlighted that regional contributions outweigh local sources in determining $\text{PM}_{2.5}$ levels, with winter months experiencing significantly higher concentrations than monsoon periods~\cite{Kumar}.

\begin{figure}[t]
    \centering   \includegraphics[width=0.6\linewidth]{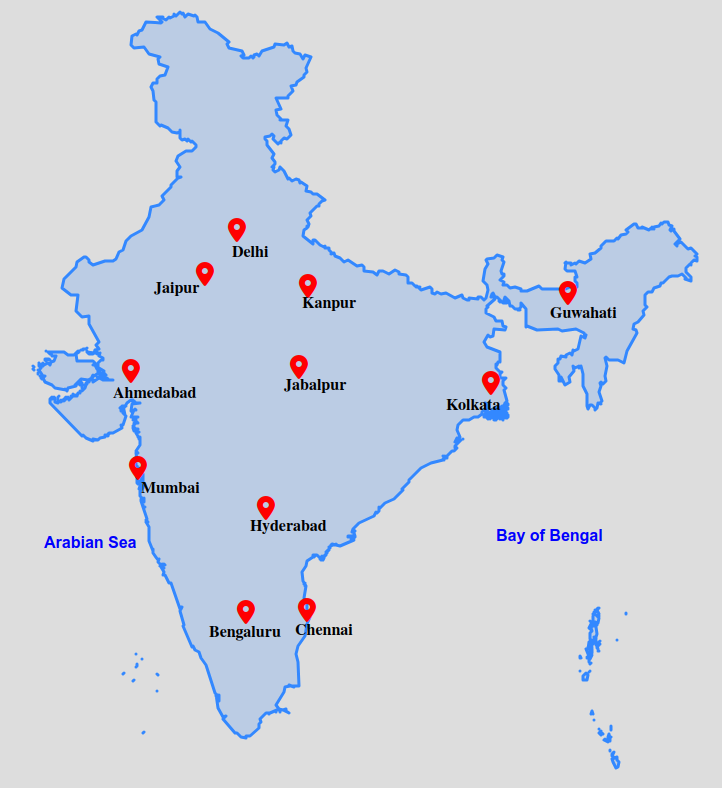}
    \caption{The Indian cities selected for this study indicated on the map.}   
    \label{fig:map}
\end{figure}

In spite of modest progress~\cite{Mishra_2021,Beck_2020} in modeling air pollutant distributions, key gaps remain in understanding the probabilistic characteristics of $\text{PM}_{2.5}$ concentrations. The probability density functions have been used~\cite{Mishra_2021} to analyze $\text{PM}_{2.5}$ pollutant data from multiple countries, with their goodness-of-fit tests using various statistical indices. 
It has also been shown~\cite{Beck_2020} that the probability distributions of certain pollutant concentrations exhibit heavy tails, with NO following a $\chi^2$ superstatistics and NO$_2$ conforming to an inverse-$\chi^2$ superstatistics. 
However, current literature largely emphasizes time-averaged trends, spatial variations, and exceedance statistics~\cite{Sree, Singh} but neglects the analysis of the randomness and inherent structure of these distributions. Notably, systematic efforts to characterize the seasonal distributions of particulate matter through advanced statistical parameters, such as entropy, and to measure their randomness remain absent. 
Furthermore, the comparison of distributional similarities between cities across different seasons using relative entropy measures remains unexplored in existing studies. However, entropy has been employed in environmental studies for predictive analysis on identifying poor air quality episodes. Some research~\cite{pred_2020,pred1_2020} integrates entropy theory with machine learning to forecast precipitation indices, demonstrating its utility in hydrological modeling.

This study primarily addresses the gaps in the relevant literature by introducing the concept of entropy.
First, it characterizes the seasonal distributions of $\text{PM}_{2.5}$ concentrations across a range of cities in India that vary in population density, pollution levels, topographical character and also in climate zones.
Our work identifies a universal exponential trend in the seasonal distributions of 
$\text{PM}_{2.5}$ concentrations across all cities, particularly at higher concentration levels. 
This trend indicates that while $\text{PM}_{2.5}$ concentrations increase at the upper end of their distributions, they do not rise indefinitely; instead, a clear exponential cutoff exists. 
Although the cutoff value varies across different regions and different seasons, the underlying pattern of an exponential decline remains consistent across all cities and seasons.  Secondly, in a crucial part of our analysis we assess the randomness of these seasonal distributions. 
We quantify the randomness using entropy, specifically Shannon entropy~\cite{Shanon}, a concept from information theory that measures the degree of uncertainty or unpredictability within a data distribution. Entropy has proven to be a valuable tool in many fields, from thermodynamics to communication theory.
Here, it serves as a measure to capture the spatiotemporal complexity of $\text{PM}_{2.5}$ concentration patterns, providing a robust framework for understanding the randomness in the distributions.

Finally, we employ the concept of relative entropy~\cite{rel_1,rel_2} to compare the seasonal probability distributions of $\text{PM}_{2.5}$ concentrations among different cities.  We adopt the Jensen-Shannon divergence (JSD)~\cite{kl,js} measure to provide a quantitative estimate of the degree of similarity between the probability distributions corresponding to two different cities.  Widely used in machine learning and information theory, JSD allows us to assess how closely the $\text{PM}_{2.5}$ concentration distributions of different cities resemble each other, which could have important implications for policy and targeted interventions.
To the best of our knowledge, this study represents one of the first efforts to use comparative analysis of $\text{PM}_{2.5}$ concentrations between various Indian cities based on their probability distributions. 
By addressing these unexplored dimensions, this work aims to enrich the understanding of $\text{PM}_{2.5}$ behavior and provide actionable insights for location and season-specific air quality management strategies to mitigate the escalating risks associated with $\text{PM}_{2.5}$ pollution in India.

The paper is organized as follows: The Data Description section provides a detailed overview of the dataset of daily $\text{PM}_{2.5}$ concentration data. 
This section also highlights the relevance of studying 
$\text{PM}_{2.5}$ distributions in these diverse regions. 
Section 3  delves into the key analysis of our results. We narrate here  the seasonal and regional variations in 
$\text{PM}_{2.5}$ distributions, their probabilistic characteristics such as entropy, and inter-city comparisons using a relative entropy measure. It also identifies a universal exponential trend in the distribution tails, indicating a bounded increase in concentrations.
Finally, in Section 4 we summarize our findings with some future directions.

\section{Data Description} 

The Central Pollution Control Board (CPCB) \url{(https://cpcb.nic.in/namp-data/)} administers a nationwide program for ambient air quality monitoring, known as the National Air Quality Monitoring Program (NAMP). 
For this study, we have selected eleven cities in India [see Fig.\ref{fig:map}], representing a mix of six megacities, with population above 10 million, and five metropolitan cities, with population between 1 and 10 million, to capture variations in geographical, climatic, and demographic attributes. 
The six megacities—{\bf Delhi, Mumbai, Kolkata, Bengaluru, Chennai, and Hyderabad}—are among the most populous and industrially significant urban areas, experiencing high levels of economic activity and diverse climatic conditions. Additionally, five metropolitan cities—{\bf Ahmedabad, Jaipur, Kanpur, Guwahati, and Jabalpur}—are included to provide a broader perspective on $\text{PM}_{2.5}$ distribution across different regions. 
These cities vary in industrialization, population density, and climatic conditions, from the arid climate of Jaipur to the humid subtropical environment of Guwahati. 
Table~\ref{tab:city} provides detailed information on the selected cities.  

\begin{table}[h]
    \centering
    \small
\resizebox{\columnwidth}{!}{%
    \begin{tabular}{|c|c|c|c|c|c|c|c|}
    \hline
     &  City name &  State/UT &  Population & Location (lat-long) 
     & Climate zone \\
     \hline 
     C1 & Delhi (DEL) & Delhi NCR &  33,807,400 & 28.61$^\circ$N, 77.21$^\circ$E 
     & Cwa, BSh\\
     \hline 
     C2 & Mumbai (MUM) & Maharashtra &  21,673,100 & 19.07$^\circ$N, 72.88$^\circ$E 
     &  Aw, Am\\
     \hline
     C3 & Kolkata (KOL) & West Bengal &  15,570,800 & 22.57$^\circ$N, 88.36$^\circ$E 
     & Aw\\
     \hline 
     C4 & Bengaluru (BLR) & Karnataka &  14,008,300 & 12.97$^\circ$N,77.59$^\circ$E 
     & Aw \\
     \hline 
     C5 & Chennai (CHE) & Tamil Nadu &  12,053,700 & 13.08$^\circ$N, 80.27$^\circ$E 
     &  As\\
     \hline 
     C6 & Hydrabad (HYD) & Telangana & 11,068,900 & 17.39$^\circ$N, 78.49$^\circ$E 
     &  BSh\\
     \hline 
     C7 & Ahmedabad (AHM) & Gujarat  & 8,854,440& 23.02$^\circ$N,72.57$^\circ$E 
     &  BSh, Aw\\
     \hline 
     C8 & Jaipur (JAI) & Rajasthan &  4,308,510 & 26.91$^\circ$N, 75.79$^\circ$E 
     & BSh \\
     \hline 
     C9 & Kanpur (KNP) & Uttar Pradesh & 3,286,140 & 26.45$^\circ$N,80.33$^\circ$E 
     & Cwa, BSh\\
     \hline 
     C10 & Jabalpur (JBP) & Madhya Pradesh  & 1,551,000 & 23.18$^\circ$N,79.95$^\circ$E 
     & Cfa\\
     \hline 
     C11 & Guwahati (GUW) & Assam &  1,199,460 & 26.14$^\circ$N, 91.74$^\circ$E 
     & Cwa, Aw\\
     \hline 
     \end{tabular}
     }
    \caption{Detailed information of the eleven selected cities. Population as of 2024 is taken from~\cite{pop} and the climate zones are according to K\"oppen's Climate Classification~\cite{Kop}.}
    \label{tab:city}
\end{table}

For this study, we collect daily $\text{PM}_{2.5}$ concentration data spanning six years, covering the period from July 1, 2018 to June 30, 2024, from all available monitoring stations in the cities C1-C9. The data for the remaining two cities C10-C11 are available only from 2019 and  we collect the data for the period  July 1, 2019 to June 30, 2024. In total, we aggregate data from 129 monitoring stations~\cite{data}.

The extended time-frame provides a robust dataset that captures seasonal and combined data variations, long-term trends, and regional differences in 
$\text{PM}_{2.5}$ concentrations across diverse cities.

\section{Results and Analysis}
\begin{figure}[t]
    \centering    \includegraphics[width=0.82\linewidth]{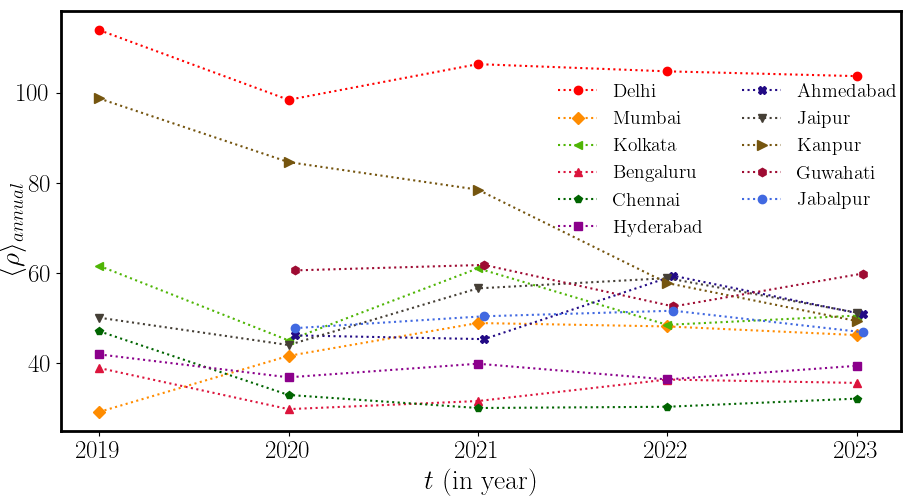}
    \caption{Trend of annual average $\la \rho \ra_{\text{annual}}$ of $\text{PM}_{2.5}$ density  for the selected eleven cities.}
    \label{fig:Annual}
\end{figure}

We start by computing the annual mean of $\text{PM}_{2.5}$ concentration $\rho$ (in $\mu g/m^{3}$). For each city, we compute the mean concentration $\langle \rho \rangle_{\text{annual}}$ over the period  January 1-December 31, for each year. Figure~\ref{fig:Annual} illustrates the temporal variation of $\langle \rho \rangle_{\text{annual}}$ for each city. Although certain cities show a slight decreasing or increasing trend  over the years, the overall annual mean concentration remains relatively stable for most cities, with no significant changes observed in the past five years. This allows us to look at the fluctuation of $\rho$ over the whole period and construct the probability density function, $P(\rho)$, of the $\text{PM}_{2.5}$ concentration for each city.  

To extract $P(\rho)$ for a city, we use the entire dataset from all the monitoring stations in that city and first construct the corresponding histogram. For each city, the bin width is set to 10 to obtain a smooth curve. The probability density function is then obtained by normalizing the histogram, so that  $\int_0^\infty d\rho P(\rho) = 1$. Note that, $P(\rho)$ has a dimension of $(\mu g/m^{3})^{-1}$.

\begin{figure}[H]
    \centering      \includegraphics[width=0.48\linewidth]{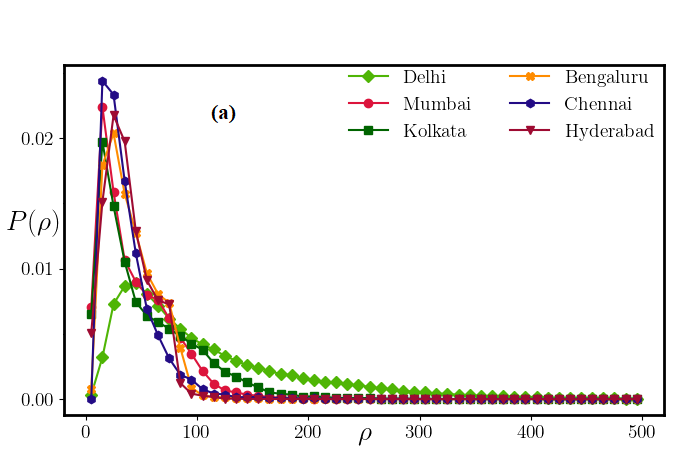}     
    \includegraphics[width=0.48 \linewidth]{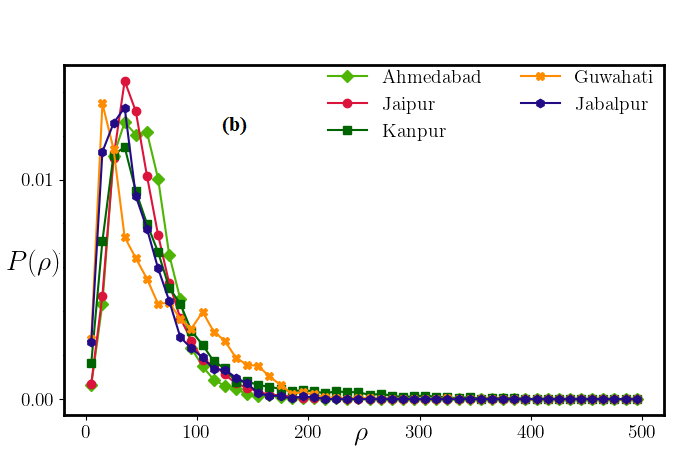}  \caption{Plot of $P(\rho)$ {\it vs} $\rho$, obtained from the aggregated data for: (a) megacities (C1-C6) and (b)  the metropolitan cities (C7-C11).}
      \label{fig:dist}
\end{figure}

The results are presented in Fig.~\ref{fig:dist}, separately for the megacities and the metropolises. Clearly, the $\text{PM}_{2.5}$ distribution has strong regional variations, although $P(\rho)$ for several of the cities show similar qualitative features. The regional variation in $\text{PM}_{2.5}$ distributions is influenced by geographical factors (e.g., coastal vs. inland cities), climatic conditions (e.g., arid, subtropical, or humid zones), and human activities (e.g., industrial emissions, transportation, and biomass burning). 
Northern inland cities like Delhi and Kanpur exhibit higher concentrations with broader distribution tails, indicating severe and persistent pollution levels.  In contrast, coastal cities like Mumbai and Chennai generally have narrower distributions and lower $\text{PM}_{2.5}$ concentrations likely due to favorable climatic conditions.

\begin{figure}[tbh]
    \centering   \includegraphics[width=0.48\linewidth]{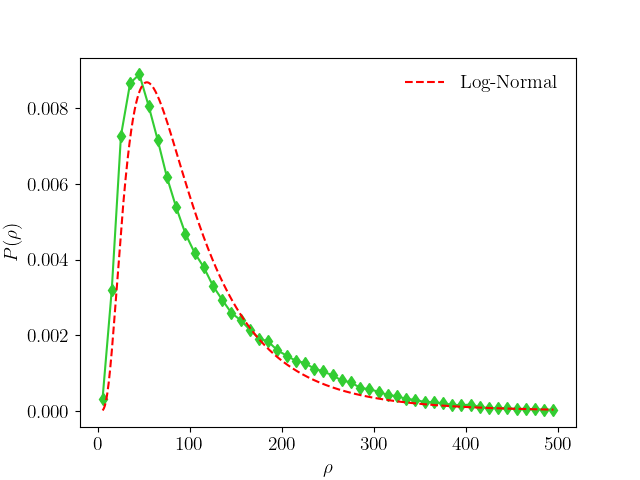} \includegraphics[width=0.48 \linewidth]{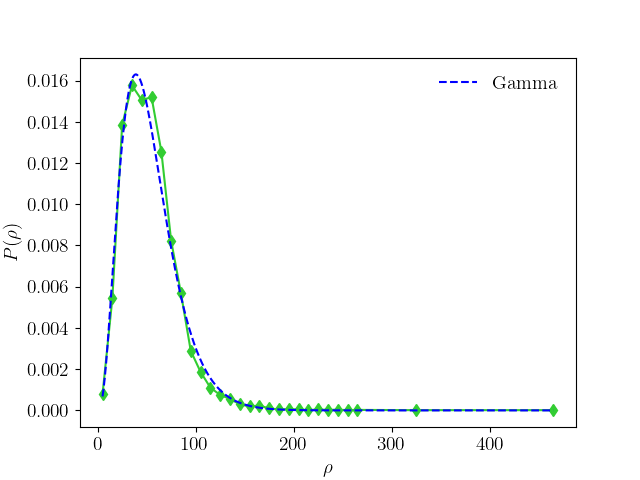}  
    \caption{Probability distributions of $\text{PM}_{2.5}$ concentrations fitted with log-normal (red dashed) and Gamma (blue dashed) curves for (a) Delhi and (b) Ahmedabad respectively.}
      \label{fig:K}
\end{figure}

It would be useful to see if the distributions follow any universal pattern across different cities. To this end, we fit the observed probability distributions to several well-established theoretical models. 
Fig.~\ref{fig:K} illustrates the best fit curves for $P(\rho)$  for two representative cities, namely, Ahmedabad and Delhi. 
In Ahmedabad, the $\text{PM}_{2.5}$ concentration data closely follows a Gamma distribution, whereas in Delhi, it aligns more closely with a log-normal distribution. Similarly, other cities exhibit diverse distributional patterns, with no single theoretical model accurately describing the data across all locations. For example, cities such as Kolkata and Guwahati exhibit a distinct double-peaked structure. Additionally, while some locations have the majority of their probability density concentrated at lower concentration levels ($\rho \le 100$), more polluted cities show substantial contributions in the tail regions, reflecting frequent occurrences of high pollution episodes. This variation highlights the absence of a universal distribution and underscores the regional differences in $\text{PM}_{2.5}$ concentration characteristics.

\begin{figure}[t]
\centering    
\includegraphics[width=0.48\textwidth]{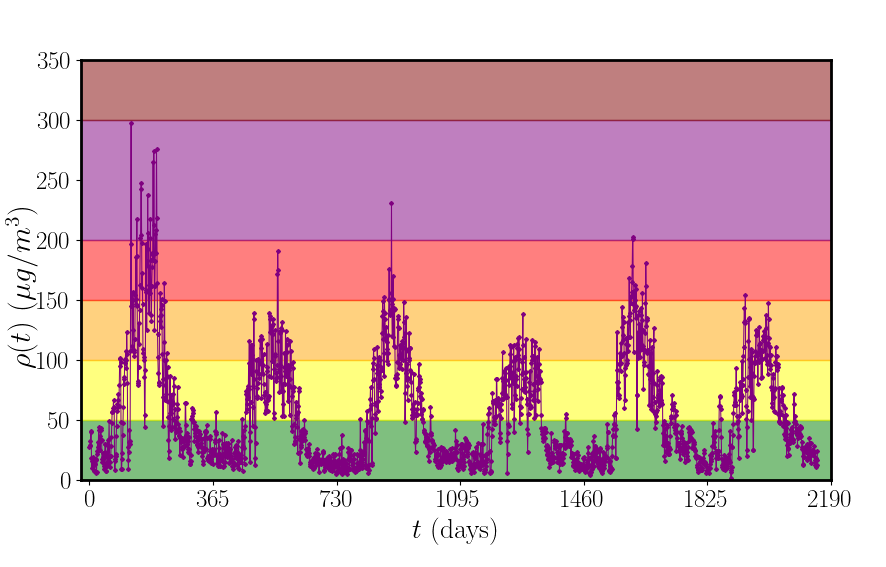}    \includegraphics[width=0.48\textwidth]{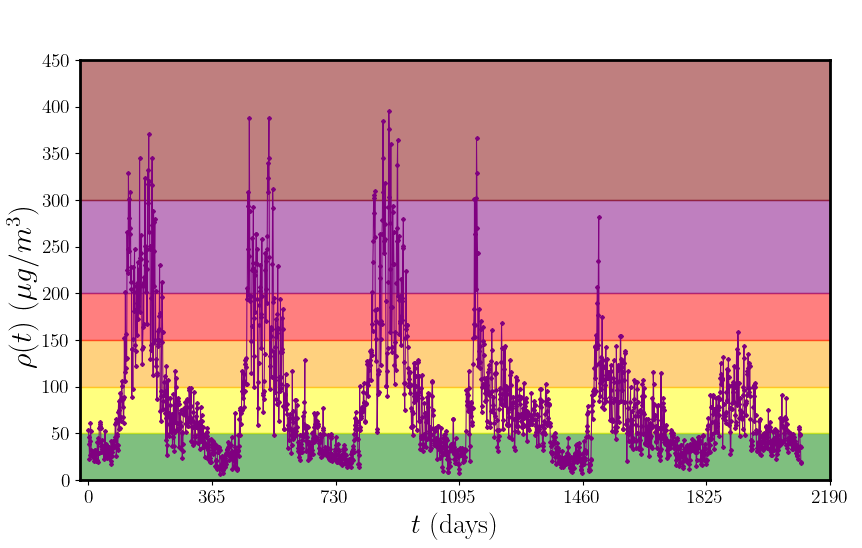}
\caption{Temporal variation of $\text{PM}_{2.5}$ from July 2018 to June 2024 for (left) Victoria, Kolkata  and (right) Nehru Nagar, Kanpur.
Each color in the background corresponds to a different level of health concern~\cite{clr}.} \label{fig:1}
\end{figure}

To gain further insights into these variations, we look at the time-series data, $\rho(t)$. Fig.~\ref{fig:1} presents the daily $\rho(t)$ plotted against time $t$ over a six-year period for two monitoring stations from two different cities.
 Both the plots display distinct periodic behavior, illustrating a recurring pattern in each year where $\text{PM}_{2.5}$ concentrations gradually rise, reach a peak during specific months, and then decline to lower levels. This cyclic pattern repeats consistently across the entire six-year period, reflecting a stable and predictable seasonal cycle. Additionally, several abrupt and sharp spikes are evident, signifying short-term pollution episodes or extreme events. For all the cities, $\rho(t)$  exhibits a consistent periodic pattern, with a distinct peak occurring annually. Recent studies~\cite{Gao_2020,Sree,Singh} also emphasize the significance of understanding seasonal variations in air pollution patterns. 

\subsection{Seasonal Variation of $\text{PM}_{2.5}$ Concentration}

India’s distinct climatic conditions, as defined by the India Meteorological Department, follow an international standard of four seasons, namely, Winter, Summer, Monsoon and Post-Monsoon with slight regional adjustments. The months representing these seasons are indicated in Table~\ref{tab:season}. Accordingly, we divide the entire dataset, for each city, into four parts. For example, the dataset for Summer for Kolkata comprises the daily data for March-May for the years 2019 to 2024.

\begin{table}[h]
    \centering
    \begin{tabular}{|p{4 cm}|p{4 cm}|}
    \hline
        \textbf{Season} & \textbf{Months} \\
        \hline
      \textbf{Winter}   &  December - February\\
      \hline
      \textbf{Summer}   &  March - May\\
      \hline
      \textbf{Monsoon}   &  June - September\\
      \hline
      \textbf{Post-Monsoon}   &  October - November\\
      \hline     
    \end{tabular}
    \caption{The months representing the four seasons.}
    \label{tab:season}
\end{table}


To characterize the seasonal variations in $\text{PM}_{2.5}$ concentration, we first measure the most common statistical parameters, namely, mean and standard deviation, for each city. For a dataset containing $N$ points $\{ \rho_i; i=1, 2, \cdots N\}$, the mean $\la \rho \ra$ and the second moment $\la \rho^2 \ra$ are given by,
\begin{align}
 \langle \rho \rangle &= \frac 1N\sum_{i=1}^N \rho_i, \quad \text{and} \quad  \langle \rho^2 \rangle =  \frac 1N\sum_{i=1}^N \rho_i^2. \label{eq:moment_def}
\end{align}
The standard deviation $\sigma$ is then computed as,
\begin{align}
 \sigma = \sqrt{\la \rho^{2} \ra  - \langle \rho \rangle ^{2}}.    \label{eq:sig_def} 
\end{align}
Note that, the size $N$ of the dataset depends on the specific season and on the number of stations available in a particular city. 

\begin{figure}[h]
     \centering    \includegraphics[width=0.65\textwidth]{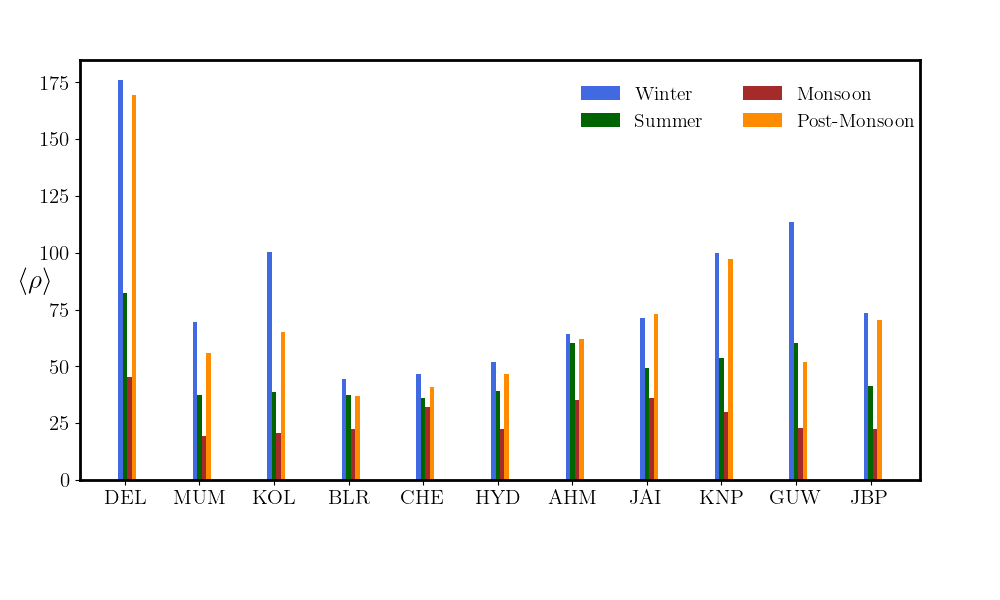}
     \caption{Mean values of each of the four seasons of $\text{PM}_{2.5}$ distributions across eleven selected cities.}
        \label{fig:mean}
\end{figure}

For each city, we compute $\la \rho \ra$ and $\sigma$ for the four seasons (according to Table~\ref{tab:season}) using Eqs.~\eqref{eq:moment_def} and \eqref{eq:sig_def}. 
The mean $\la \rho \ra$ provides a fundamental measure of the central tendency, reflecting the average pollution levels experienced in each city. This parameter highlights long-term exposure trends and helps identify cities with consistently higher or lower particulate matter concentrations. On the other hand, the standard deviation $\sigma$ measures the variability or dispersion of $\text{PM}_{2.5}$ concentrations around the mean. A higher
standard deviation indicates greater fluctuations in pollution levels, suggesting the influence of episodic events or seasonal changes that drive significant deviations from the average concentration.
\begin{table}[h]
 \centering
 \small
\resizebox{\columnwidth}{!}{%
\begin{tabular}{|c|c||c|c|c|c|c|c|c|c|}
\hline
 & &\multicolumn{2}{c|} {Winter} & \multicolumn{2}{c|} {Summer}  & \multicolumn{2}{c|} {Monsoon} & \multicolumn{2}{c|} {Post-Monsoon} \\
 \cmidrule(lr){3-10}
& \shortstack[lb]{Cities} & $\langle \rho \rangle$ & $\sigma$ & $\langle \rho \rangle $ & $\sigma$ & $\langle \rho \rangle $ & $\sigma$ & $\langle \rho \rangle $ & $\sigma$ \\
\hline
C1 & DEL & 176.06 & 85.96 & 82.39 & 39.47 & 45.37 & 28.98 & 169.35 & 108.16\\
\hline
C2 & MUM & 69.63 & 34.80 & 37.40 & 26.36 & 19.15 & 24.79 & 55.95 & 27.70\\
\hline
C3 & KOL & 100.42 & 43.40 & 38.63 & 22.46 & 20.75 & 12.09 & 65.15 & 38.64\\
 \hline
C4 & BLR & 44.45 & 36.28 & 37.27 & 33.51 & 22.40 & 28.29 & 37.14 & 29.84\\
\hline
C5 & CHE & 46.77 & 33.56 & 36.31 & 38.60 & 31.94 & 36.10 & 41.09 &	31.96\\
\hline
C6 & HYD & 51.92 & 22.10 & 39.31 & 22.01 & 22.63 & 12.96 & 46.72 & 21.76\\
\hline
C7 & AHM & 64.30 & 27.30 & 60.37 & 26.64 & 35.04 & 22.63 & 61.94 & 29.03\\
\hline
C8 & JAI & 71.42 & 30.95 & 49.48 & 24.09 & 36.01 & 21.61 & 73.17 & 36.45\\
\hline
C9 & KNP & 99.71 & 67.37 & 53.55 & 22.99 & 29.75 & 16.04 & 97.17 & 64.56\\
\hline 
C10 & JBP & 73.36 & 34.42 & 41.42 & 17.29 & 22.38 & 11.18 & 70.29 & 34.67\\
\hline
C11 & GUW & 113.55 & 38.78 & 60.30 & 46.40 & 23.07 & 16.88 & 52.14 & 25.40\\
\hline
\end{tabular}
 }
\caption{Mean and standard deviation of distributions of $\text{PM}_{2.5}$ concentration for each of the four seasons.}
\label{tab:3}
\end{table}
Table~\ref{tab:3}  provides a comprehensive summary of the $\langle \rho \rangle$ and $\sigma$  for each season for all the cities. The seasonal variation of $\la \rho \ra$ is also depicted 
in Figure~\ref{fig:mean} for the different cities. This analysis reveals certain generic features of the seasonal fluctuations of $\text{PM}_{2.5}$.
We have noticed that, for all the cities, $\text{PM}_{2.5}$ concentrations are typically at their highest during winter season, decrease in summer, reach their lowest levels during the monsoon, and rise again in the post-monsoon period. Recent research~\cite{Gao_2020} attributes elevated winter pollution to increased emissions from heating activities, higher vehicular and industrial output, and reduced atmospheric dispersion caused by temperature inversions and lower temperatures. Summer brings improved air quality due to enhanced atmospheric mixing, stronger winds, and higher temperatures that aid in dispersing pollutants, although emissions from transportation and industry remain significant. Monsoon rains act as an effective cleansing agent, removing particulate matter from the atmosphere and resulting in the lowest pollution levels. However, in the post-monsoon period, $\text{PM}{2.5}$ concentrations rise again, likely due to the resumption of intensified emission activities and unfavorable meteorological conditions, including reduced rainfall and cooling temperatures that hinder pollutant dispersion. Notably, regional climate variations cause some cities to experience peak pollution levels in the post-monsoon season.

\subsubsection{Seasonal Distributions }
To understand the seasonal variations of $\text{PM}_{2.5}$ concentration fluctuations in more details, it is imperative to look at its distributions for each season separately. To this end,  we first construct the seasonal probability distribution of the $\text{PM}_{2.5}$ concentrations for each city, using seasonal datasets. Unless otherwise specified, we use a bin width 10, which is optimized to ensure smooth distributions. Note that, for notational convenience we denote all these distributions using the same letter $P$.

\begin{figure}[h]
     \centering
\includegraphics[width=0.4\textwidth]{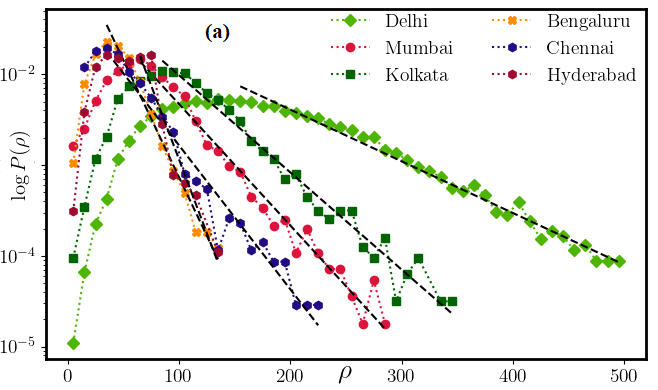}
\includegraphics[width=0.4\textwidth]{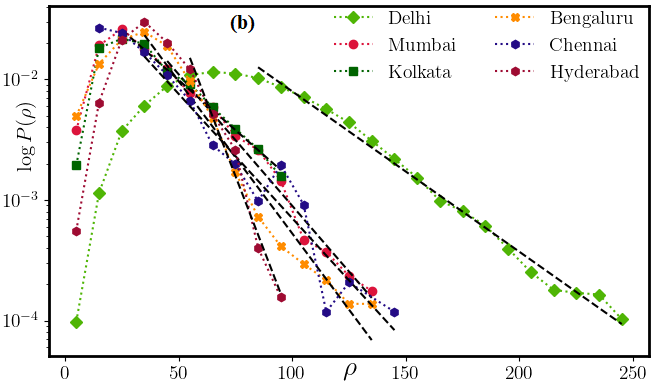}
\includegraphics[width=0.4\textwidth]{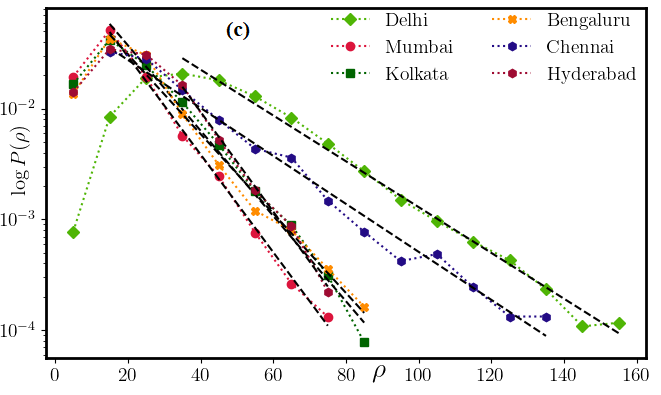}
\includegraphics[width=0.4\textwidth]{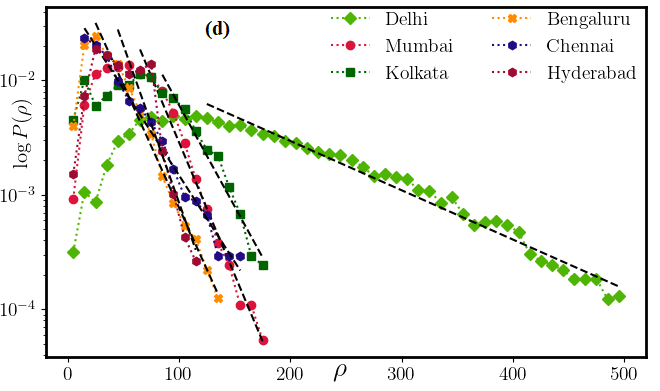}
\caption{Exponential tail fitting of seasonal $\text{PM}_{2.5}$ distributions for the cities C1-C6: (a) Winter, (b) Summer, (c) Monsoon, and (d) Post-Monsoon.} 
 \label{fig:Expo_mega}
\end{figure}

\begin{figure}[h]
     \centering
\includegraphics[width=0.4\textwidth]{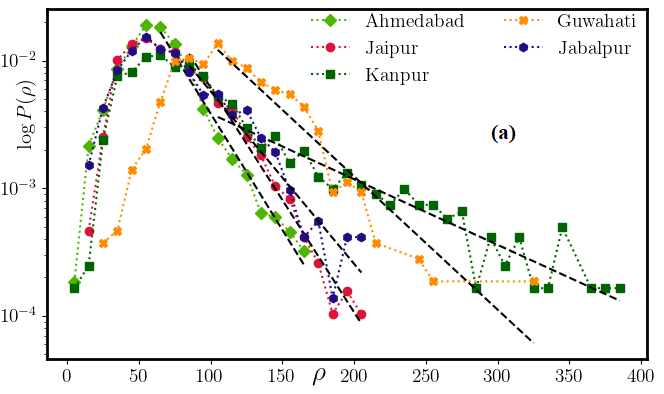}
\includegraphics[width=0.4\textwidth]{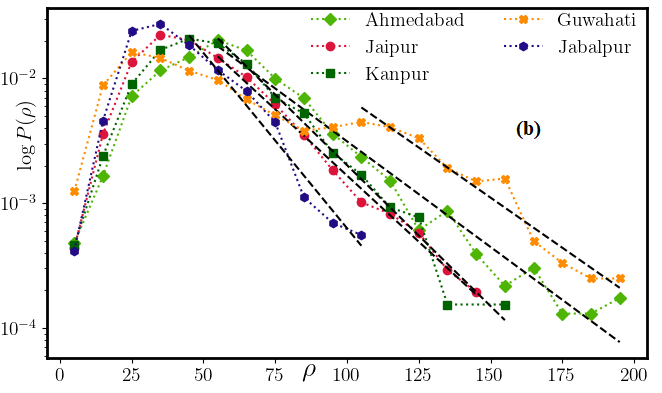}
\includegraphics[width=0.4\textwidth]{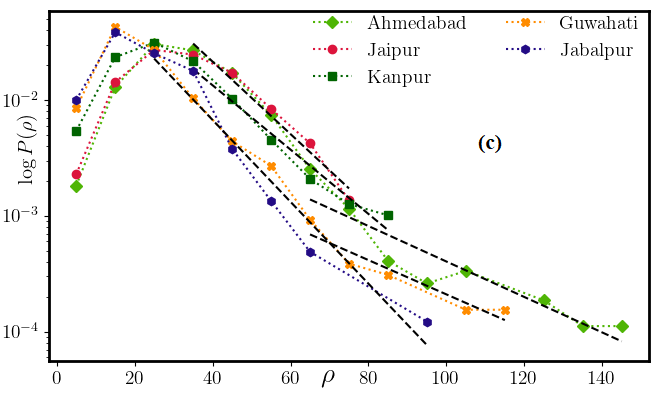}
\includegraphics[width=0.4\textwidth]{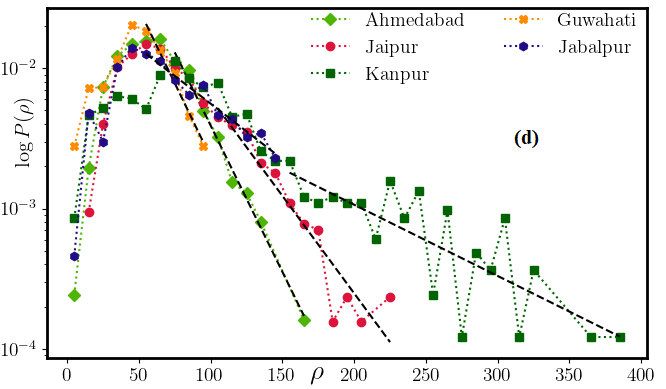}
\caption{Exponential tail fitting of seasonal $\text{PM}_{2.5}$ distributions for the cities C7-C11: (a) Winter, (b) Summer, (c) Monsoon, and (d) Post-Monsoon. }
 \label{fig:Expo_other}
\end{figure}

Figures~\ref{fig:Expo_mega} and \ref{fig:Expo_other} show plots of the seasonal $P(\rho)$ for all the cities in semi-log scale.
Interestingly, our analysis reveals a universal feature in $\text{PM}_{2.5}$ concentrations, where the tail regions of $P(\rho)$ consistently follow an exponential decay. However, the decay exponents vary across cities and seasons, highlighting regional and temporal variations in pollution patterns. To estimate these decay exponents, we fit the tails of these distributions to a functional form,
\begin{equation}
P(\rho) \sim \exp {\left(- \rho/ \theta \right)},
\end{equation}
where the exponent $\theta$ provides the typical scale for the $\text{PM}_{2.5}$ concentration level. 
The values of $\theta$ obtained from this fitting are summarized in Table~\ref{tab:1}. For instance, cities with consistently high $\text{PM}_{2.5}$ levels, such as Delhi or Kanpur, show a slower exponential decay, reflecting a higher likelihood of extreme pollution events. In contrast, cities like Bengaluru or Chennai, with relatively lower pollution levels, exhibit steeper decay rates, suggesting that extreme $\text{PM}_{2.5}$ events are less frequent. 
As observed in Table~\ref{tab:1}, the seasonal variations of $\theta$ across different cities confirm that during the winter season, $\text{PM}_{2.5}$ concentrations reach their peak levels, while in the summer, they begin to decrease. The concentrations reach their lowest levels during the monsoon season, followed by a gradual increase in the post-monsoon season. 
\begin{table}[h]
 \centering
\begin{tabular}{|c|c|c|c|c|c|}
\hline
& Cities & Winter & Summer  & Monsoon & Post-Monsoon \\
\hline
C1 & DEL & $76.92 \pm 0.0003$ & $32.26 \pm 0.0007$ & $20.83 \pm 0.0013$ & $101.01 \pm 0.0003$ \\
\hline
C2 & MUM & $32.47 \pm 0.0012$ & $20.41 \pm 0.0025$ & $9.80 \pm 0.0025$ & $18.76 \pm 0.0021$ \\
 \hline
C3 & KOL & $40.65 \pm 0.0009$ & $24.39 \pm 0.0017$ & $11.24 \pm 0.0039$ & $24.39 \pm 0.0029$ \\
\hline
C4 & BLR & $16.95 \pm 0.0029$ & $17.24 \pm 0.0039$ & $12.05 \pm 0.0038$ & $20.16 \pm 0.0015$\\
\hline
C5 & CHE & $27.78 \pm 0.0017$ & $21.28 \pm 0.0047$ & $20.02 \pm 0.0019$ & $28.57 \pm 0.0011$\\
\hline
C6 & HYD & $13.33 \pm 0.0091$ & $8.85 \pm 0.0110$ & $9.62 \pm 0.0048$ & $15.87 \pm 0.0099$ \\
\hline
C7 & AHM & $23.81 \pm 0.0020$ & $25.64 \pm 0.0022$ & $28.57 \pm 0.0057$ & $20.83 \pm 0.0019$ \\
\hline
C8 & JAI & $24.63 \pm 0.0019$ & $20.83 \pm 0.0014$ & $13.89 \pm 0.0077$ & $31.25 \pm 0.0034$\\
\hline
C9 & KNP & $84.75 \pm 0.0008$ & $19.23 \pm 0.0033$ & $15.63 \pm 0.0055$ & $83.33 \pm 0.0021$\\
\hline
C10 & JBP & $32.05 \pm 0.0035$ & $15.38 \pm 0.0058$ & $12.29 \pm 0.0093$ & $54.95 \pm 0.0014$\\
\hline
C11 & GUW & $41.67 \pm 0.0023$ & $27.03 \pm 0.0030$ & $29.41 \pm 0.0065$ & $20.58 \pm 0.0042$\\
\hline 
\end{tabular}
\caption{Estimated values of $\theta$ with corresponding errors for each of the four seasons across the eleven selected cities.}
\label{tab:1}
\end{table}

This universal exponential decay is one of the most important results of this work. This implies that despite differences in geographical locations, climatic conditions, and anthropogenic activities, the extreme values of $\text{PM}_{2.5}$ concentrations (representing the tail of the distributions) follow a consistent exponential behavior across all cities. 
It also suggests that mitigation strategies for extreme pollution can be widely applicable, with localized adjustments based on regional exponent values.

\subsubsection{Shannon Entropy of City-wise Seasonal distributions}
To further characterize  the randomness of the seasonal distributions, we also compute the Shanon entropy, a fundamental information theoretic measure~\cite{Shanon}. Shannon entropy  measures the degree of uncertainty or diversity in a probability distribution. For a continuous distribution, it is defined as,
\begin{equation}
 S = -\int_{0}^{\infty} d \rho \, P(\rho)\ln (P(\rho)).  \label{eq:S_def}
\end{equation}
This metric $S$, which can take any real value, provides a robust means of assessing how evenly the $\text{PM}_{2.5}$ concentrations are distributed across regions, helping to identify patterns in the data that might not be apparent through traditional statistical measures like moments alone.
Higher entropy values suggest a more uniform spread of concentration levels across the observed range, while lower entropy points to a comparatively narrow distribution with concentration of values around specific levels, such as consistent high or low pollution episodes. 

\begin{table}[h]
 \centering
\begin{tabular}{|c|c||c|c|c|c|}
\hline
& Cities & Winter & Summer  & Monsoon & Post-Monsoon \\
\hline
C1 & DEL & 5.750 & 4.997 & 4.456 & 5.868 \\
\hline
C2 & MUM & 4.891 & 4.408 & 3.537 & 4.689 \\
\hline
C3 & KOL & 5.102 & 4.360 & 3.694 & 4.950 \\
 \hline
C4 & BLR & 4.354 & 4.257 & 3.721 & 4.392 \\
\hline
C5 & CHE & 4.295 & 3.588 & 3.405 & 4.499 \\
\hline
C6 & HYD & 4.394 & 4.150 & 3.605 & 4.432 \\
\hline
C7 & AHM &	4.679 & 4.627 & 4.097 & 4.643 \\
\hline
C8 & JAI & 4.597 & 4.441 & 4.218 & 4.941 \\
\hline
C9 & KNP & 5.304 & 4.421 & 4.043 & 5.380 \\
\hline 
C10 & JBP & 4.785 & 4.200 & 3.696 & 4.880 \\
\hline
C11 & GUW & 5.106 & 4.909 & 4.022 & 4.562 \\
\hline
\end{tabular}
\caption{Entropy values of distributions of $\text{PM}_{2.5}$ concentration for each of the four seasons across the eleven selected cities.}
\label{tab:4}
\end{table}
\begin{figure}[h]
     \centering     \includegraphics[width=0.6\linewidth]{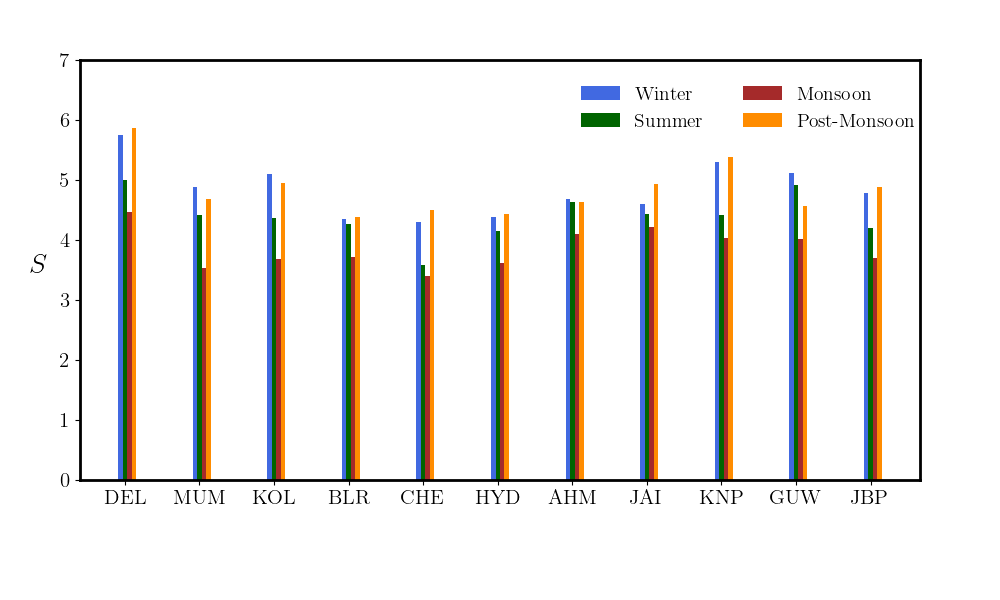}
     \caption{Entropy values of seasonal and combined data of $\text{PM}_{2.5}$ distributions across eleven selected cities.}
    \label{fig:ent}
\end{figure}

For each city, we compute $S$ for all the four seasons.
These are summarized in Table~\ref{tab:4}. Figure~\ref{fig:ent} also shows a bar plot of seasonal $S$ for the different cities.
Clearly, winter exhibits overall higher entropy values, with a maximum of 5.750 in Delhi and a minimum of 4.295 in Chennai, indicating a broader spread of pollution concentrations. 
Conversely, monsoon features reduced entropy values, with a peak of 4.456 in Delhi and a low of 3.405 in Chennai, indicating narrower distributions likely influenced by increased rainfall and washout effects. 

Computation of the Shannon entropy allows us to classify the cities according to the level of randomness in their seasonal $\text{PM}_{2.5}$ fluctuations. In particular, we focus on winter and monsoon, the two seasons with typically highest and lowest entropy values. For each of the two seasons, we classify the cities in four groups according to decreasing values of entropy, namely,  extremely high, high, medium and low entropy groups. This classification is summarized in Table~\ref{tab:entropy}.
\begin{table}[h]
\centering
\small
\resizebox{\columnwidth}{!}{%
\renewcommand{\arraystretch}{1.5}
\begin{tabular}{|c|c|c|}
\hline
\textbf{Season} & \textbf{Class} & \textbf{Cities} \\ 
\hline
\multirow{4}{*}{\textbf{Winter}} 
& \textbf{Extremely High Entropy} & Delhi, Kanpur  \\ 
\cmidrule{2-3} 
& \textbf{High Entropy} & Kolkata, Guwahati  \\ 
\cmidrule{2-3} 
& \textbf{Medium Entropy} & Ahmedabad, Jaipur, Jabalpur, Mumbai \\ 
\cmidrule{2-3}  
& \textbf{Low Entropy} & Bengaluru, Chennai, Hyderabad \\ 
\hline
\hline
\multirow{4}{*}{\textbf{Monsoon}} 
& \textbf{Extremely High Entropy} & Delhi\\ 
\cmidrule{2-3} 
& \textbf{High Entropy} & Ahmedabad, Jaipur, Guwahati, Kanpur \\ 
\cmidrule{2-3} 
& \textbf{Medium Entropy} & Kolkata, Jabalpur, Bengaluru, Hyderabad \\ 
\cmidrule{2-3} 
& \textbf{Low Entropy} & Mumbai, Chennai\\ 
\hline
\end{tabular}
}
\caption{Classification of cities based on entropy of $\text{PM}_{2.5}$ distributions during winter and monsoon seasons.}
\label{tab:entropy}
\end{table}

The observation aligns well with seasonal mean and standard deviation trends [in Table~\ref{tab:3}], reinforcing the robustness of this approach. This grouping correlates with climate zones, where arid or inland regions demonstrate broader distributions and higher seasonal variations, while humid or coastal areas benefit from more stable and predictable air quality patterns. Such consistency between entropy-based classification and statistical measures enhances the understanding of pollution dynamics and facilitates more targeted air quality management strategies.

Further, classifying cities according to entropy values provides a valuable framework that can be further utilized for predictive analysis~\cite{Ana_2020}. Instead of analyzing each city in isolation, this classification captures shared distributional characteristics and variability patterns, enabling more efficient and generalized forecasting models of air quality patterns and pollution events.

\subsection{Intercity Comparative Study of Seasonal $\text{PM}_{2.5}$ Distributions Through Relative Entropy}

In this section we present a comparative analysis of the seasonal $\text{PM}_{2.5}$ distributions of different cities using a relative entropic measure, namely,  the Jensen-Shannon divergence. The Jensen-Shannon divergence (JSD)~\cite{js} between two probability distributions $P_{i}(\rho)$ and $P_{j}(\rho)$ is defined as,
\begin{equation}
    D_{ij}(P_{i}||P_{j}) = \frac{1}{2} \int d \rho ~ \Bigg [P_i(\rho) \ln \frac{P_i (\rho )}{M_{ij}(\rho)} + P_j(\rho) \ln \frac{P_j (\rho )}{M_{ij}(\rho)} \Bigg],
 \end{equation}
 where $M_{ij}(\rho) = \frac{1}{2}[P_{i}(\rho)+P_{j}(\rho)]$ denotes the average of $P_{i}(\rho)$ and $P_{j}(\rho)$. JSD is nothing but a symmetrized version of Kullback-Leibler divergence~\cite{kl} and  provides a quantitative measure of the degree of similarity between two distributions. Clearly, the JSD for two identical  distributions vanishes. In fact, it is easy to show that JSD is always non-negative and is bounded in the regime $D_{ij} \in  [0, \ln 2 \simeq 0.693]$ with a lower value indicating a higher degree of similarity. 

In this work, we measure $D_{ij}$ for each pair of cities $(Ci,Cj)$ where $P_{i}(\rho)$ and $P_j(\rho)$ denote their corresponding $\text{PM}_{2.5}$  distributions in a given season. Once again, we focus on the two extreme seasons, namely, winter and monsoon. The resulting matrices are presented in Tables~\ref{tab:JS} and \ref{tab:JS1}, respectively. A close examination of these results reveals that for both the seasons, the JSD has very low values for a set of pair of cities. This, in turn, implies that,  within a season, some cities have highly similar distributions. To identify such groups, we apply a supervised learning algorithm, k-means clustering~\cite{k_means_1967,k_means_1982}, to the $D_{ij}$ values for winter and monsoon, separately. This analysis helps us identify a set of closely lying $D_{ij}$ values which are substantially lower than the rest. For winter, these low JSD values correspond to the range $D_{ij} \le 0.05$ while for monsoon, we get $D_{ij} \le 0.025$.

Let us consider the scenario for the winter first. From the Table~\ref{tab:JS}, it is evident that the values of $D_{ij} < 0.05$ naturally divide into two distinct sub-classes with no values lying within the range $[0.026, 0.038]$. The first sub-class with $D_{ij}  \in [0, 0.026]$ indicates a high degree of similarity, while the second sub-class with  $D_{ij} \in [0.038,0.050]$, suggests moderately similar distributions.  These subclasses are indicated  by blue and red shaded cells in Table~\ref{tab:JS} where the former denotes a pair of cities with  high distributional similarity and the latter signifies moderately similar distributions.

\begin{table}[h]
    \centering
 \small
\resizebox{\columnwidth}{!}{%
    \begin{tabular}{|c|c|c|c|c|c|c|c|c|c|c|c|}
    \hline
       &  \shortstack[lb]{DEL\\(C1)} & \shortstack[lb]{MUM\\(C2)} & \shortstack[lb]{KOL\\(C3)} & \shortstack[lb]{BLR\\(C4)} & \shortstack[lb]{CHE\\(C5)} & \shortstack[lb]{HYD\\(C6)} & \shortstack[lb]{AHM\\(C7)} & \shortstack[lb]{JAI\\(C8)} & \shortstack[lb]{KNP\\(C9)} & \shortstack[lb]{JBP\\(C10)} & \shortstack[lb]{GUW\\(C11)}\\
     \hline 										 \shortstack[lb]{DEL\\(C1)} &  &  0.305 &	0.155 &	0.503 & 0.429 &	0.476 &	0.364 &	0.298 &	0.155 &	0.292 &	0.129\\
     \hline 
     \shortstack[lb]{MUM\\(C2)}  &  &  & 0.083 &	0.131 &	0.088 &	0.077 &	\cellcolor{blue!25} 0.019 &	\cellcolor{blue!25} 0.004 &	\cellcolor{red!25} 0.049 &	\cellcolor{blue!25} 0.011 &	0.188\\
     \hline 
     \shortstack[lb]{KOL\\(C3)}  &  &  &  & 0.330 &	0.256 &	0.261 & 0.136 & 0.064 &	\cellcolor{red!25} 0.045 & 0.064 &	\cellcolor{red!25} 0.038\\
     \hline
     \shortstack[lb]{BLR\\(C4)}  &   &  &  & &\cellcolor{blue!25}  0.012 &\cellcolor{red!25}	0.049 & 0.113 &	0.147 &	0.219 &	0.148 & 0.466\\
     \hline
     \shortstack[lb]{CHE\\(C5)}  &  &  &  &  &  & \cellcolor{red!25} 0.042 & 0.086 &	0.111 &	0.171 &	0.108 &	0.382\\
     \hline
     \shortstack[lb]{HYD\\(C6)} &  & &  &  & & &  \cellcolor{red!25} 0.050 &	0.094 &	0.169 &	0.099 &	0.392\\
     \hline
     \shortstack[lb]{AHM\\(C7)}  &  & &  &  &  &  &  &  \cellcolor{blue!25} 0.013 &	0.078 &	\cellcolor{blue!25} 0.026 &	0.257\\ 
     \hline
     \shortstack[lb]{JAI\\(C8)} &  &  &  & &  & & & & \cellcolor{red!25} 0.039 & \cellcolor{blue!25} 0.006 &	0.168\\
     \hline
     \shortstack[lb]{KNP\\(C9)}  &  & & &  &  &  &  &  &   & \cellcolor{red!25} 0.048 &	0.122\\
     \hline
     \shortstack[lb]{JBP\\(C10)} &  & & &  &  &  &  &  &  & &  0.160 \\
     \hline
     \shortstack[lb]{GUW\\(C11)}  &  & & &  & &  &  & & & & \\
     \hline
   \end{tabular}
   }
    \caption{Tabulation form of the matrix of Jensen-Shannon Divergence (JSD) values for the winter season. The blue colored cell ($D_{ij} \in [0,0.026]$) and red colored cell ($D_{ij} \in [0.038,0.050]$) represent the high similarity and the moderate similarity between cities, respectively.}
    \label{tab:JS}
\end{table}

For example, Bengaluru and Chennai exhibit a high degree of similarity in  their $\text{PM}_{2.5}$ distributions, forming a distinct pair. Additionally, Mumbai, Ahmedabad, Jaipur, and Jabalpur cluster together, displaying strong distributional resemblance. Interestingly, these groupings are consistent with entropy-based classifications [see Table~\ref{tab:entropy}]. Mumbai, Ahmedabad, Jaipur, and Jabalpur not only share similar distribution patterns but also exhibit comparable levels of randomness in their $\text{PM}_{2.5}$ concentrations, forming a well-defined group. Similarly, Bengaluru and Chennai align closely in both distributional similarity and entropy measures, while Hyderabad, exhibiting nearly the same level of randomness with these cities, displays moderately similar distributional patterns with them.  Delhi, on the other hand, stands out as highly distinct from all other cities, indicating that its distributional characteristics differ markedly. Furthermore, Kolkata's distributional pattern is moderately similar to that of Guwahati and Kanpur, while Kanpur shares moderate similarity with Jaipur, Jabalpur, and Mumbai in winter season. 

\begin{figure}[h]
    \centering
    \includegraphics[width=0.48\linewidth]{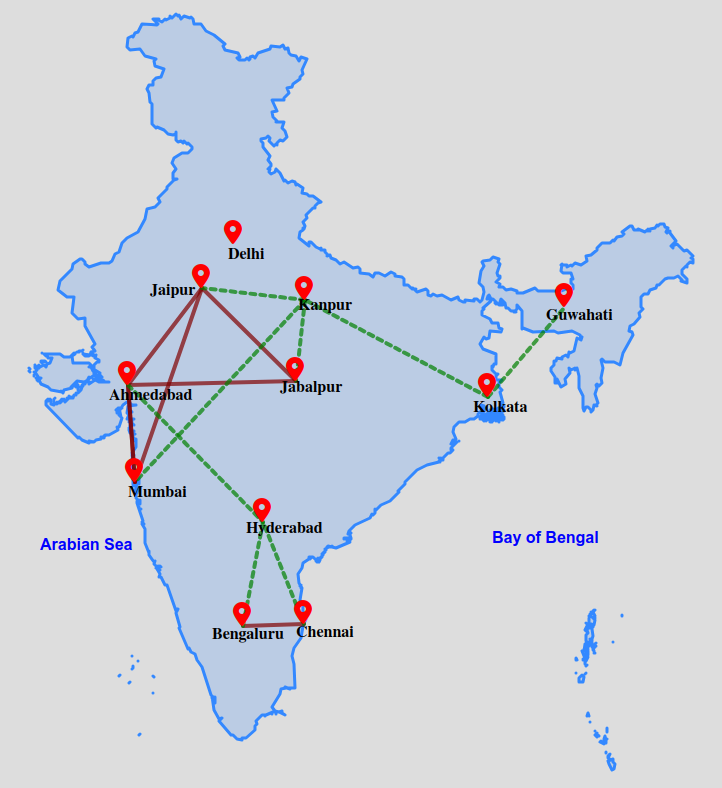}
    \includegraphics[width=0.48\linewidth]{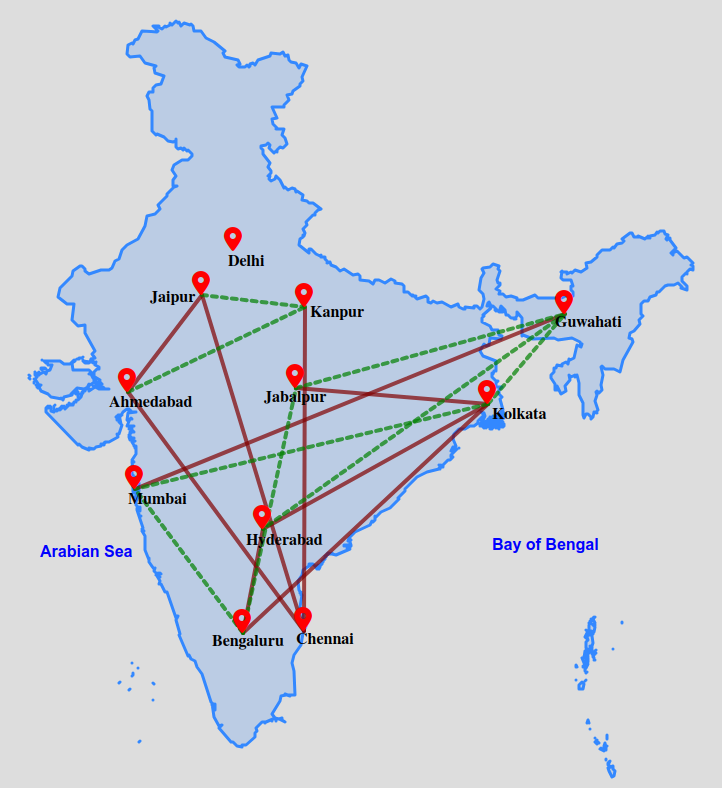}
    \caption{Intercity comparisons of $\text{PM}_{2.5}$ distributional patterns in Winter (left) and Monsoon (right) season. Solid red lines indicate cities with very similar distributions, while dashed green lines represent moderately similar distributions.}
    \label{fig:map_connect}
\end{figure}

This grouping of cities based on their $\text{PM}_{2.5}$ distributional similarities is displayed on a map of India in the left panel of Fig.~\ref{fig:map_connect}. 
In the figure, cities that exhibit highly similar $\text{PM}_{2.5}$ distributions are interconnected by solid red lines, indicating a strong resemblance in their pollution patterns. Meanwhile, cities with moderately similar distributions are linked by dashed green lines, signifying a noticeable but less pronounced similarity in their $\text{PM}_{2.5}$ concentration distributions. By visually mapping the geographic locations of cities, this representation reveals spatial grouping trends and potential regional influences that shape air pollution patterns. 


Next, we examine the scenario for the monsoon season when $\text{PM}_{2.5}$ concentrations reach their lowest levels.  Following the same procedure as before, we identify city groups with highly similar ($D_{ij }< 0.011$) and moderately similar ($ 0.017 \le D_{ij} \le 0.025$) distributions [see Table~\ref{tab:JS1}]. The overall reduction in pollution levels during monsoon results in the many of the cities having similar distributions in this season. Consequently, we get a much larger group, illustrated on the right panel of Fig.~\ref{fig:map_connect}. Interestingly, Delhi still stands out with a distinct distributional pattern, differing from all other cities. A comparison of the two panels in Fig.~\ref{fig:map_connect} reveals how seasonal variations affect the degree of similarity between cities across different regions of India.

\begin{table}[t]
    \centering
 \small
\resizebox{\columnwidth}{!}{%
    \begin{tabular}{|c|c|c|c|c|c|c|c|c|c|c|c|}
    \hline
       &  \shortstack[lb]{DEL\\(C1)} & \shortstack[lb]{MUM\\(C2)} & \shortstack[lb]{KOL\\(C3)} & \shortstack[lb]{BLR\\(C4)} & \shortstack[lb]{CHE\\(C5)} & \shortstack[lb]{HYD\\(C6)} & \shortstack[lb]{AHM\\(C7)} & \shortstack[lb]{JAI\\(C8)} & \shortstack[lb]{KNP\\(C9)} & \shortstack[lb]{JBP\\(C10)} & \shortstack[lb]{GUW\\(C11)}\\
     \hline 							
     \shortstack[lb]{DEL\\(C1)} &  & 0.275 & 0.200 &	0.215 &	0.041 &	0.196 & 0.039 &	0.031 &	0.074 &	0.187 &	0.206\\
     \hline 
     \shortstack[lb]{MUM\\(C2)}  &  &  & \cellcolor{red!25} 0.020 &	\cellcolor{red!25} 0.019 &	0.131 &	0.101 &	0.198 &	0.194 &	0.108 &	0.038 &	\cellcolor{blue!25} 0.010\\
     \hline 
     \shortstack[lb]{KOL\\(C3)}  &  &  &  & \cellcolor{blue!25} 0.008 &	0.072 & \cellcolor{blue!25}	0.001 & 0.121 &	0.117 &	0.050 &	\cellcolor{blue!25} 0.008 &	\cellcolor{red!25} 0.017\\
     \hline
     \shortstack[lb]{BLR\\(C4)}  &   &  &  & & 0.065 &	\cellcolor{blue!25} 0.011 & 0.138 &	0.132 &	0.062 &	\cellcolor{red!25} 0.017 &	0.051\\
     \hline
     \shortstack[lb]{CHE\\(C5)}  &  &  &  &  &  & 0.051 &	\cellcolor{blue!25} 0.005 &	\cellcolor{blue!25} 0.006 &	\cellcolor{blue!25} 0.001 &	0.070 &	0.098\\
     \hline
     \shortstack[lb]{HYD\\(C6)} &  & &  &  & & &  0.087 &	0.100 &	0.058 &	0.033 &	\cellcolor{red!25} 0.023\\
     \hline
     \shortstack[lb]{AHM\\(C7)}  &  & &  &  &  &  &  &  \cellcolor{blue!25} 0.005 &	\cellcolor{red!25} 0.022 &	0.101 &	0.187\\ 
     \hline
     \shortstack[lb]{JAI\\(C8)} &  &  &  & &  & & & & \cellcolor{red!25} 0.022 &	0.100 &	0.121\\
     \hline
     \shortstack[lb]{KNP\\(C9)}  &  & & &  &  &  &  &  &   & 0.038 &	0.051\\
     \hline
     \shortstack[lb]{JBP\\(C10)} &  & & &  &  &  &  &  &  & &  \cellcolor{red!25} 0.025 \\
     \hline
     \shortstack[lb]{GUW\\(C11)}  &  & & &  & &  &  & & & & \\
     \hline
   \end{tabular}
    }
    \caption{Tabulation form of the matrix of Jensen-Shannon Divergence (JSD) values for the monsoon season. The blue colored cell ($D_{ij} \in [0,0.011]$) and red colored cell ($D_{ij} \in [0.017,0.025]$) represent the high similarity and the moderate similarity between cities, respectively.}
    \label{tab:JS1}
\end{table}

\section{Concluding Remarks}
Controlling air pollution, particularly fine particulate matter ($\text{PM}_{2.5}$), remains a critical environmental and public health challenge in India. 
This study provides a comprehensive analysis of $\text{PM}_{2.5}$ concentrations across eleven Indian cities, spanning diverse geographical and climatic zones as well as population range.  
This work uniquely contributes by delving into the probabilistic structure, randomness, and inter-city comparisons of seasonal and regional $\text{PM}_{2.5}$ distributions.

Considering the daily $\text{PM}_{2.5}$ concentration for over six years (2018-2024) and characterizing its distributions, we identify a universal exponential tail behavior across all the cities in all the four seasons, though with distinct decay rates reflecting regional and seasonal differences. Such tail behavior is critical in understanding extreme pollution events that significantly impact health and quality of life. 

The analysis of the information-theoretic measure, entropy, provides a comprehensive framework for characterizing the seasonal variations in $\text{PM}_{2.5}$ distributions. Entropy ($S$) serves as a valuable tool for quantifying the complexity and randomness within these distributions, offering insights into the variability of air pollution across different cities. By examining entropy values, we can systematically classify cities into distinct groups based on their distributional randomness during winter and monsoon. Cities that belong to the same group share similar levels of randomness in their $\text{PM}_{2.5}$ distributions, suggesting that they experience comparable pollution patterns. Notably, this classification is further supported by statistical parameters such as the mean and standard deviation, which exhibit consistent patterns with the identified groups.

Moreover, a key contribution of this study is the use of Jensen-Shannon divergence (JSD), a relative entropy-based metric, to assess the similarities in $\text{PM}_{2.5}$ concentration distributions across various cities. By applying this novel approach, the study provides a deeper understanding of the complex regional and inter-city characteristics of air pollution. 
This comprehensive analysis uncovers distinct groupings of cities during the winter and monsoon seasons, shaped by their varying degrees of distributional similarity. These findings highlight the value of advanced comparative metrics, such as JSD, in crafting tailored, region-specific strategies that account for seasonal variations in urban air pollution patterns across India.

In conclusion, this work advances the understanding of air pollution in India by introducing innovative statistical and information-theoretic approaches to characterize and compare $\text{PM}_{2.5}$ distributions. The integration of entropy and relative entropy measures provides a robust framework for assessing the variability and inter-regional relationships of air pollution. These findings not only strengthen our understanding of air quality dynamics but also offer actionable insights for designing region-specific, evidence-based strategies to combat air pollution in India. 
Moreover, the identification of the consistent groups with entropy as well as relative entropy values in winter season is instrumental for policymakers aiming to design effective strategies for mitigating $\text{PM}_{2.5}$ pollution. Rather than addressing air pollution at the individual city level, these groups enable a collaborative, region-focused approach to tackling this issue. Such coordination among cities within the same group can significantly enhance the efficiency of mitigation efforts, reduce costs, save time, and contribute to the overall reduction of national pollution levels.
This framework ensures a more impactful and cohesive response to one of the most pressing environmental challenges.

It would be interesting to see whether the same spatio-temporal pattern is observed for a broader range of pollutants such as $\text{SO}_2$, $\text{NO}_2$ etc.
Additionally, exploring the influence of climate variables—such as temperature, humidity, and wind patterns—on $\text{PM}_{2.5}$ distributions across different climatic zones could yield valuable insights. Recent studies~\cite{aditya_2018,samad_2023} demonstrate the effectiveness of machine learning techniques, particularly decision tree algorithms~\cite{pred_2023,d-tree}, for predicting air pollutant concentrations. Integrating entropy as a decision-making tool within these models could provide a novel approach to enhancing the seasonal predictability of air quality across different cities. The findings of this study offer a foundation for future predictive frameworks that improve air pollution forecasting and inform proactive management strategies.


\bibliography{ref}

\end{document}